\documentclass[conference,a4paper]{IEEEtran}

\usepackage[style=ieee,doi=true,url=false]{biblatex}
\bibliography{jasmin}

\ifCLASSINFOpdf
 \usepackage[pdftex]{graphicx}
\else
  \usepackage[dvips]{graphicx}
\fi
\usepackage{fixltx2e}

\usepackage{stfloats}

\usepackage{url}

\usepackage{booktabs}

\begin{document}
%
\title{The JASMIN super-data-cluster}

\author{\IEEEauthorblockN{
B.N. Lawrence\IEEEauthorrefmark{1}\IEEEauthorrefmark{2}\IEEEauthorrefmark{4}, 
V. Bennett\IEEEauthorrefmark{2}\IEEEauthorrefmark{5}, 
J. Churchill\IEEEauthorrefmark{3}, 
M. Juckes\IEEEauthorrefmark{2}\IEEEauthorrefmark{4}, 
P. Kershaw\IEEEauthorrefmark{2}\IEEEauthorrefmark{5}, 
P. Oliver\IEEEauthorrefmark{3},
M. Pritchard\IEEEauthorrefmark{2}\IEEEauthorrefmark{4}\IEEEauthorrefmark{5}}
\IEEEauthorblockN{
and 
A. Stephens\IEEEauthorrefmark{2}\IEEEauthorrefmark{4}}
\IEEEauthorblockA{
\IEEEauthorrefmark{1}Department of Meteorology, University of Reading, Reading, U.K.
}
\IEEEauthorblockA{
\IEEEauthorrefmark{2}Centre for Environmental Data Archival, STFC Rutherford Appleton Laboratory, Didcot, U.K.}
\IEEEauthorblockA{
\IEEEauthorrefmark{3}E-Science Department, STFC Rutherford Appleton Laboratory, Didcot, U.K.}
\IEEEauthorblockA{
\IEEEauthorrefmark{4}National Centre for Atmospheric Science
\IEEEauthorrefmark{5}National Centre for Earth Observation}

}


\maketitle                        

\begin{abstract}
The JASMIN super-data-cluster is being deployed to support the data analysis 
requirements  of the UK and European climate and earth system modelling
community. Physical co-location of the core JASMIN resource with significant 
components of  the facility for Climate and Environmental Monitoring from Space (CEMS) 
provides additional support for the earth observation  community, 
as well as facilitating further comparison and evaluation of models
with data. JASMIN and CEMS together centrally deploy 9.3 PB of storage --- 4.6 PB of Panasas fast disk storage 
alongside the STFC Atlas Tape Store.
Over 370 computing cores provide local
computation. Remote JASMIN resources at Bristol,
Leeds and Reading provide additional distributed storage and compute configured to support local workflow as a stepping stone to using the central JASMIN system. Fast
network links from JASMIN provide reliable communication between
the UK supercomputers MONSooN (at the Met Office) and HECToR (at the University
of Edinburgh). JASMIN also supports European users via a light path to KNMI
in the Netherlands. The functional components of the JASMIN infrastructure have 
been designed to support and integrate workflows for three main goals: (1) the efficient operation of data
curation and facilitation at the STFC Centre for Environmental Data Archival;
(2) efficient data analysis by the UK and European climate and earth system
science communities, and; 
(3) flexible access for the climate impacts and earth observation 
communities to  complex data and concomitant services.
\end{abstract}

\vspace{1em}


\section{Introduction}

JASMIN, the Joint Analysis System Meeting Infrastructure Needs, is a super-data-cluster, deploying petascale fast disk connected via low latency networks to significant amounts of data-analysis compute. It is deployed in the e-Science department at the STFC Rutherford Appleton Laboratory (RAL) on behalf of the National Centre for Atmospheric Science (NCAS, one of six research centres of the UK Natural Environment Research Council, NERC). It is closely coupled to infrastructure deployed on behalf of the NERC Centre for Earth Observation (NCEO).

The JASMIN compute and network environment will be managed by the STFC e-Science department enabling the scientific functionality to be delivered by staff from the STFC Centre for Environmental Data Archival (CEDA, see \url{http://www.ceda.ac.uk}).  

In this paper we  introduce the rationale for the deployment of this facility in terms of the scientific need and the infrastructural context (national and international data and compute resources) and then describe the system itself and some of the workflow we expect to see deployed on JASMIN.

It will be seen that JASMIN serves to deliver three main functions:  the infrastructure for the data storage and services of CEDA; an environment for data intensive scientific computation for the climate and earth system science communities; and flexible access to high volume and complex data as well as processing services for the climate and earth observation communities. 
    
\subsection{Scientific Requirement}

Much of our understanding of the climate system is encoded in complex numerical models. Such models have been constructed to work on range of scales - from large eddy simulations aimed at understanding the physics and dynamics of cloud systems, to full earth system models aimed at understanding the system over millennia. Whatever the aim of such models, whether they simulate phenomena or systems, or are for understanding, prediction, or projection, they need to be evaluated both by intercomparison and confrontation with data.

The scale of the data handling required has become intimidating: for example, the fifth global climate model intercomparison project (CMIP5) is expected to produced three petabytes (PB) of "requested" output (that is, output which has been identified as especially useful for intercomparison). The requested data is being catalogued and made available globally via software developed by the Earth System Grid Federation (ESGF) and deployed by the Global Organisation for Earth System Science Portals. The structure and running of this global federation is  complicated, e.g. \cite{Williams_earth_2011}, and not discussed in detail here.  As well as the global federated requested data, there will probably be tens of PB of additional data produced which may or may not appear outside the originating institutions using the federated system. Along with the model data, significant amounts of observational data are being made available for intercomparison with the model simulations. In the U.S.A, NASA/JPL are leading the way by reformatting and exposing earth observation data using the same ESGF tools. In Europe, ESA is sponsoring the Climate Change Initiative (the "CCI").

While storing and disseminating the bits-and-bytes at these volumes is not trivial, it is not the volume alone that makes the data handling intimidating.  There are millions of files, many of which offer ostensibly the same data products, for the same time, with the differences due to subtleties in model initialisation and/or configuration e.g. \cite{guilyardi_cmip5_2011}. All these lead to major issues for diverse user communities distributed geographically and across scientific disciplines, with very different technical skills, workflows and software ecosystems.

Even where communities share a close scientific heritage, and the same basic skillset, handling these datasets efficiently is a problem - especially when collaborative work is necessary. For example, a workshop held in 2010 on joint research activities between the Met Office and the NERC earth system modelling community highlighted the lack of a joint analysis facility as a significant hurdle for collaborative model development and the consequential scientific exploitation. This echoed similar concerns echoed within the NERC community itself many times over the last decade - even within NERC there are communities contributing to common earth system modelling efforts who are both geographically disparate and who do not share common analysis facilities. It will be seen that one of the major objectives of JASMIN is to address this requirement of a shared analysis environment.

The increasing volume and quality of climate and earth observation data is leading to the proliferation of new user communities.  Since the previous major model intercomparison project (CMIP3), thousands of new users from communities hitherto either uninterested in such data, or unable to use them, have engaged with the data and data services. (Examples of such users include the insurance community and environmental consultants.) However, many are still significantly inhibited by lack of access to appropriate computing resources (whether local or remote). "Appropriate" in this context includes suitable for handling and processing high volume data, and critically, deployed with suitable and familiar software.  

Of course many potential users of such data do not need access to the raw data, generally they are interested in small subsets, customised maps, or statistics. There are a number of ways of addressing such users which do not involve abandoning them to sink or swim with complicated-to-use computing resources. The two most effective are to establish collaborative teams which include sophisticated users experienced with the data, and to deploy customised data portals; it will be seen that JASMIN will support both techniques. However, while both are effective, both have drawbacks: the first does not scale well to big or diverse user communities, and the second cannot anticipate all possible combinations of required data. Hence there is a requirement to provide a computing ecosystem which supports very diverse computing environments, portals, and as many as possible of the nuts and bolts of portal services. Again, JASMIN is configured to support this diversity as far as possible, and CEMS will provide an intermediate solution which involves users being able to upload and execute code in customised environments.

Looking to the future, the problems with data handling for both earth observation and climate and earth system science are only going to increase as data volumes increase exponentially \cite{overpeck_climate_2011}.  From a data analysis point of view the problems come from multiple directions: we now have multiple HPC systems generating data  (as will be seen below), and we have an increased diversity of observational data sources required to evaluate models (as models encapsulate new processes). Even without the expected transition to exascale computing later this decade, increases in model complexity and resolution, coupled with massive increases in ensemble size will lead to vast increases in the amount of data expected to be acquired and analysed. (It should be noted that while increases in model complexity and resolution may require heroic efforts to increase model scaling, larger ensemble sizes come easily from the embarrassingly parallel nature of that part of the problem.) Of course future earth observation sensors are coming with higher spatial, temporal and spectral resolution, and concomitant data volume increases as well. These problems are well recognised in the community and funding agencies.

\subsection {Infrastructural Context}

The UK academic climate modelling community have access to a range of computing facilities, from local clusters through national facilities to international access both via informal and formal relationships such as PRACE \footnote{PRACE: the Partnership for Advanced Computing in Europe, see \url{http://www.prace-ri.eu}}. Not surprisingly then, simulations are performed wherever suitable resources are available, and generally brought home to user institutions if possible. However, as outlined above, many projects are now producing too much data to be fully analysed at many home institutions - but most of the available computing environments in the UK do not have enough disk and/or archive to hold the simulations long enough for subsequent analysis. The lack of storage is exacerbated by the need to compare model simulations with each other and with data and compounded by the available network bandwidth. 

Collaboration further complicates the situation. For example, the UPSCALE joint project between the Met Office and NCAS, is exploiting the PRACE supercomputer HERMIT, to run climate models at very high resolution. 200 TB of data are to be produced in a year, and the current plan expects archival both by NCAS and the Met Office (the two partners). At least in this case there are only 2 data transfers involved. With larger collaborations, and multiple sites involved (say $N$ computational sites and $M$ analysis sites), one can anticipate up to $N \times M$ data transfers, with attendant storage problems. JASMIN is designed to mitigate against this situation - by providing enough reliable storage and compute, entire datasets need only be moved once (from wherever they are produced).

As outlined above, storage and compute are not the only problem though. Software matters too, and often partners in large collaborations have very different analysis environments. As a consequence, JASMIN has been configured to provide an environment where virtual machines can be configured by individual groups within teams to provide appropriate analysis environments - without the JASMIN system administration team having to handle each and every software environment and license\footnote{However,software licensing in private virtual clouds is not exactly a solved problem, with both vendors and buyers still finding their way in complex environments - and appropriate license vocabularies and standards not yet in place - e.g. \url{http://www.dmtf.org/news/pr/2011/2/dmtf-examine-need-software-license-management-standards}}.

There are four specific institutional contexts which are reflected in the configuration of JASMIN and associated networks:

{\bf Distributed National "Centres"}: Both NCAS and NCEO are distributed centres with units embedded primarily in universities. Some of those units are large, and include skilled support staff, but most are small. In terms of climate research, the largest groups are in Reading and Leeds, but major efforts exist in Edinburgh and Bristol (the latter not directly affiliated to either NCAS or NCEO). Climate models are also in use in much smaller research groups.  NCAS and NCEO also include data centres (themselves embedded in STFC and providing the core of CEDA) which are required to both curate research data products {\it and} use their skills and data to facilitate research in the centres and the wider community.

{\bf The Joint Weather and Climate Research Programme(JWCRP)}:This joint activity between the UK Met Office and the wider NERC community (including oceanography, land surface, etc.) has two major elements, the first of which includes a joint commitment to sustaining and growing (amongst other things) the UK's modelling capability and relevant infrastructure. The second is to align major research initiatives to ensure the most effective impact of that research (and consequential pull-through into services and policy. 

{\bf The International Space Innovation Centre (ISIC)}: is a not for profit company set up in 2010 as a partnership of universities, industry and the public sector. The ISIC objective is to generate economic growth from developing and exploiting space data and technologies. One of the ISIC activities is the facility for climate and environmental monitoring from space (CEMS) which aims to provide national data services to the scientific, government and commercial user communities. STFC, NCEO, and CEDA (via both NCEO and STFC) are all partners in ISIC and CEMS.

{\bf European Shared Infrastructure}: STFC is a partner in the European Network for Earth System Modelling (ENES, http://www.enes.org). ENES is working towards shared infrastructure to support common goals, and one manifestation of that is a collaboration between the Royal Dutch Meteorological Institute (KNMI) and BADC which aims to make the BADC data archives transparently available to staff at the KNMI and Wageningen University via a SURFnet sponsored lightpath between CEDA and those institutes.

\subsection{The role of CEDA}

CEDA currently hosts four formal data centres: the NCAS British Atmospheric Data Centre (BADC), the NCEO NERC Earth Observation Data Centre (NEODC), the IPCC Data Distribution Centre, and the UK Solar System Data Centre, as well as small research programmes in atmospheric science and data curation technologies. 

NEODC holds approximately 0.3 PB of earth observation data - much of which has, or could have, utility both in evaluating models or in supporting environmental monitoring from space.  BADC also has current holdings of around 0.3 PB, with major expansion expected as the community makes a step change in modelling resolution. For example, the CORDEX regional climate model intercomparison programme, being supported by BADC, is expected to involve another 0.3 PB of data. A number of other similar scale projects are expected to start in the next year or so. However, despite these big activities, over the next two years or so, BADC data volumes are likely to be dominated by CMIP5 support - because BADC is one of three centres worldwide who have committed to attempt to take a complete copy of the CMIP5 requested data. BADC has made this commitment in part because CEDA provides key components of the IPCC Data Distribution Centre under contract to the UK government, and it is intended that the IPCC data centres keep a tagged copy of the CMIP5 data used in their fifth assessment report. Apart from CMIP5 data, the IPCC and solar system data volumes are minimal in comparison to the NEODC and BADC.

The four CEDA data centres have been delivered by a compendium of computing solutions that have built up over the years based on network attached storage linked to an array of physical and virtual servers which support the various views into the data that are required by the data centres and their specific communities.  As CEDA has moved from terascale to petascale in the last year or so, the problems of physically managing data in this heterogeneous environment with physical devices (e.g. NAS boxes with 100 TB file systems) which were small relative to the size of the archive have begun to introduce significant operational problems. One of the gaols of JASMIN is to alleviate this situation.

Many data centres provide high volume data archives by means of relatively small disk caches coupled to large tape archives holding most of the data. CEDA does not - a copy of all data is held on disk, with tape used only as backup. This is because the current usage pattern is for users to select files from within datasets and given the diversity of data and the near random access pattern, there is no obvious way to provide an efficient tape archive. 

Some CEDA users can then further process their data by subsetting and/or visualising data using web services, but most users download data. Selecting and downloading files from datasets rather than downloading entire datasets reduces the volumes of data that users need to transfer. However, rapidly growing data volumes are making both these and straight downloading unusable with the existing technology. Users need more flexible services and the ability to do a range of calculations on the data without moving it offsite.  Again, JASMIN is intended to support such activity. 

\section{JASMIN}

In this section we describe the physical infrastructure of the JASMIN super-data-cluster and how it is configured and related to key UK and European computing infrastructure relevant to climate science.

JASMIN consists of one core system --- the JASMIN super-data-cluster --- and three satellite systems at Bristol, Leeds and Reading universities. Each of the satellite systems consists of significant disk (150, 100, and 500 TB respectively) and compute resources.

\begin{figure}[htp]\centering
\includegraphics[scale=0.7]{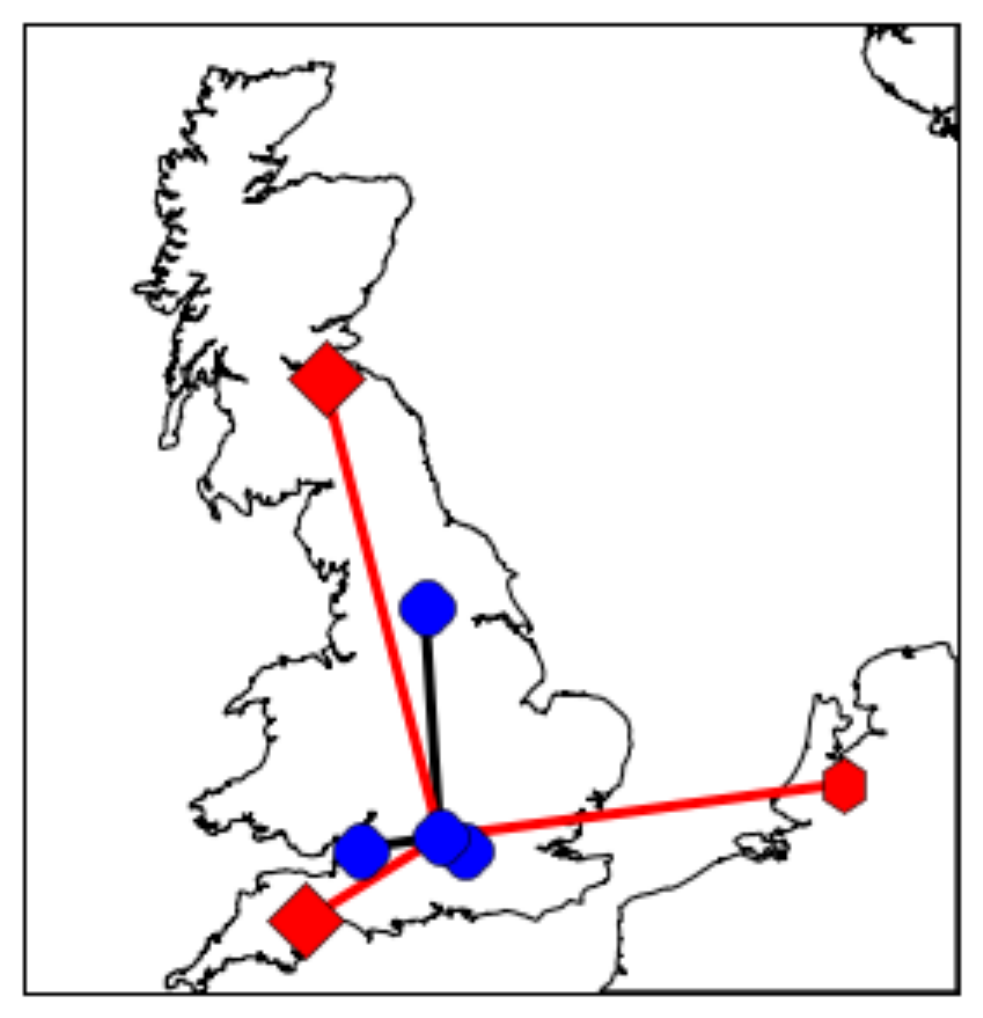}
\caption{JASMIN structure and links: The four JASMIN sites (Leeds, Bristol, Reading and RAL), linked by production JANET networks, are shown with blue circles. Remote sites linked by light paths (the national tertiary storage system and super computer in Edinburgh, the Met Office in Exeter, and KNMI in the Netherlands) are shown with red diamonds.}
\label{fig:geojas}
\end{figure}

The storage component of JASMIN is tightly integrated with a significant component of the storage required by CEMS - in practice much of the same data is required by both communities. However, CEMS have an additional component of storage of a similar scale to the Reading component of JASMIN so that "commercial" data can be held separately from the "academic" data held in JASMIN/CEMS.  This extra storage, and a significant amount of compute, is physically located in a different (ISIC) building, allowing the commercial partners to develop different levels of service quality if they so desire - however it is connected to the main JASMIN-CEMS infrastructure by a 10 Gbit/s network link. ISIC commercial services to third parties will be provided over a logically distinct network from the academic network. (Although it may be provisioned using the same physical network,  this arrangement allows JANET to charge appropriately where the network is being used commercially.)

\subsection{The JASMIN Core System}

The central JASMIN facility consists of five components:
\begin{enumerate}
\item  The core is a ring of four low latency Gnodal GS4008 switches configured as a single stack, each with 40 10 Gbit/s ports for the system, and connected bidirectionally with four 40 Gbit/s links each way to its neighbours providing a 320 Gbit/s stack backbone. The network is
configured in a number of vlans:\begin{itemize}
\item A 10 Gbit shared data vlan for JASMIN and CEMS 
\item A 10 Gbit private iSCSI vlan for VMware machine images,
\item A 10 Gbit HPC vlan for "Lotus", the JASMIN HPC cluster.
\item A 1 Gbit shared VMware server management Vlan,
\item A 10/100 Mbit Management network
\end{itemize}
\item The main JASMIN fast storage, 6.6 PB of Panasas storage, configured to provide 4.6 PB of usable storage, with significant fault tolerance via multiple levels of parity and redundancy.
\item  The main JASMIN data compute system consisting of 12 Dell R610 servers, each configured with a 12 core 3 GHZ Xeon X5675 and 96 GB of memory.
\item One Dell 815 server is attached to the network, configured with 48 2.3 GHz AMD cores (Opteron 6276) and 256 GB of memory, to provide the core data intensive web interface and provide compute for data management tasks such as housekeeping checksums etc. (It is unlikely that this will be sufficient in the long-run, but some of the legacy compute from the existing CEDA environment may be pooled with this during the transition discussed below).
\item JASMIN also includes a small HPC compute cluster, "Lotus", consisting of 8 more R610's, but with faster (3.5 GHz) processors, and less memory (48 GB). For these systems the second network link is onto a private network to support MPI traffic between nodes.
\end{enumerate}
The Panasas storage will be deployed in multiple "bladesets", that is, the storage will be subdivided into groups of storage blades that operate autonomously. Bladesets are being used for three reasons:
\begin{enumerate}
\item Fault tolerance:although there is a lot of fault tolerance in the default configuration, blade failures will happen. If one happens within a bladeset, only the failed blade will need rebuilding, and the bladeset will remain online. In what is expected to be the exceedingly rare occasion of two happening in the same bladeset at the same time, bladesets will have to rebuild offline. Other bladesets could remain available.
\item Integrity: JASMIN will distinguish between data which is being curated, and that which is in an earlier
phase of its lifecycle. Although not required by the Panasas system, by putting such data on its own
bladeset, human policies and procedures can be put in place to mitigate against inadvertent data corruption etc. We also map bladesets onto whole equipment racks powered by single electrical power phases. This tends to mitigate the effects of losing a single electrical phase, disabling a whole bladeset rather than portions of several bladesets, while allowing other bladesets to remain available.
\item  Governance: the system described here serves multiple communities, and has multiple income streams. The use of bladesets can support the human transparency between funding and resource allocation.
\end{enumerate}

\begin{figure}[htb]\centering
\includegraphics[scale=0.47]{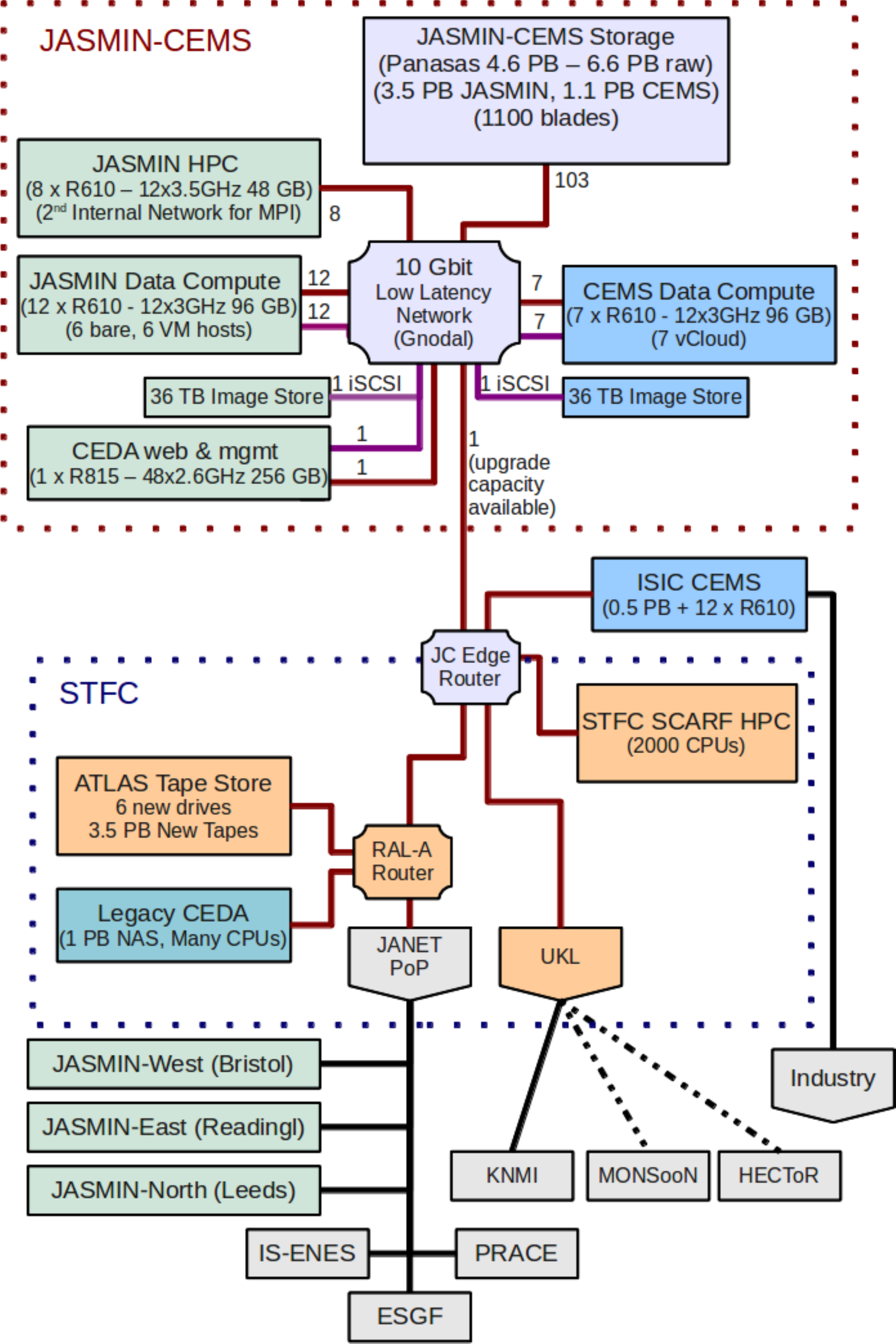}
\caption{Key JASMIN components: there are three major sections to JASMIN: the core facilities, other STFC facilities, and the distributed infrastructure - including the remote JASMIN components. The core facilities include the fast storage (connected into the low latency network with 103 10-Gbit/s connections), the JASMIN and CEMS data compute services, the JASMIN HPC, JASMIN and CEMS image stores, and the CEDA management compute. Important STFC facilities include the Atlas Tape Store and the legacy CEDA computing environment. Further CEMS compute is available at the RAL site in the ISIC building. Remote networks into JASMIN currently include light paths to KNMI (in the Netherlands) via the UK-Light router and connections to the rest of the world via the JANET RAL point of presence. Future light paths to MONSooN and HECToR are being procured.}\label{fig:Jasmin}
\end{figure}
The use of bladesets determines performance. In principle the storage provides an aggregated bandwidth of 1 Tbit/s - significantly more than the compute can handle. However, any one file can only be stored on one bladeset, so it is interesting to ask what the maximum performance single file I/O might see? Without going into details of Panasas configuration (see for example, \cite{welch_scalable_2008}), the smallest useful unit to bundle into a bladeset is  a "shelf" - with about 10 storage blades, 45 TB of storage, and a 10 Gbit/s storage network interface. The initial JASMIN configuration has three 8-shelf bladesets (with the rest of the storage
unallocated) and so if a large file was fully distributed across the bladeset, it can be accessed at a maximum
of 80 Gbit/s - the same as the maximum I/O performance of the Lotus JASMIN HPC cluster. Performance scales with bladeset size, so with the largest bladeset currently envisaged of 1 PB, or about 22 shelves,
an aggregate maximum performance of 220 Gbit/s can be achieved. This can support both Lotus and all the "non-CEMS" compute accessing it simultaneously. The
exact configuration of the bladesets when the system is fully configured will be a function of requirements, experience, and load.

The JASMIN data compute is initially configured to operate with half the servers operating as physical compute systems and half as virtual machine (VM) hosts - each with two 10 Gbit network links, one for data access, and one for access to the JASMIN 36 TB Dell Equallogic image store for the VM's (each operating on a different vlan, so as to isolate the VM image I/O traffic from the uncontrolled user data access).  The physical compute systems are intended to be used for parallelised data access. The JASMIN VM hosts will initially either be configured manually by CEDA staff to support specific user communities or be manually copied into the JASMIN environment from one of the remote JASMIN sites in Bristol, Leeds or Reading.

The long term balance between the number of machines in each configuration is not known, so there is considerable flexibility in the configuration - if more machines are needed in the physical pool, machines can be removed from the VM host pool, and vice versa. If necessary machines could also be migrated from the HPC cluster into the data server pool.

Also connected to the main network is a set of 7 CEMS data compute servers, configured with the same hardware as the JASMIN compute servers and having their own private VM image store. These servers are also intended to be run as VM hosts, but unlike the JASMIN VM hosts, it is intended that the virtual machines will be commissioned both locally and both remotely {\it and} automatically - with the resource partitioned between the local and remote usage.  It is possible that in the longer term some or all the JASMIN virtual machines could also be deployed in this manner.

\subsection {The JASMIN environment at STFC}

JASMIN is connected to the wider STFC network via an edge router, itself directly connected to the rest of CEMS in the ISIC building, the UK-Light router, the STFC HPC cluster ("SCARF"), and RAL router-A. As well as providing the default network route to the rest of the world via the JANET point of presence, Router-A provides links into two other important STFC resources: the legacy CEDA storage and compute, and the Atlas Petabyte Storage System (tape robots and front-ends).

The legacy CEDA compute resource will be maintained until all the data is migrated into JASMIN, and all the compute servers are either retired, re-purposed, or themselves migrated into the JASMIN fabric. The data transition will not be trivial - over one PB of data will need to be migrated, with contingent requirements on altering relevant metadata and ensuring data integrity during the transfer.

\subsection {Storage and Backup}

The volume of Panasas purchased was a convolution of budget (of course) and requirements. Table \ref {tab:storage} shows how the overall storage is initially expected to be used  - although actual practice may vary. 400 TB of CEDA secondary data (duplicate copies held on disk) will not be migrated onto spinning disk because the original reasons for holding this data on disk (supporting parallel access and fast recovery from errors for key data) are expected to be delivered via the increased performance and reliability of the Panasas storage.

The Atlas Data Store is maintained by the STFC e-Science department to provide scientific data services for a range of communities. In the past CEDA has used the Atlas store to provide tape backup only with all the CEDA services provided directly from NAS disk.  To support the initial backup function the Atlas Data Store has been provisioned with an additional 3.5 PB of tape to cover the JASMIN data disk (While NEODC has some tape provision, until new tapes are procured, new NEODC and CEMS backup will initially have to use the Panasas storage).  CEDA already has 1.3 PB on tape, so at launch, JASMIN has 4.8 PB of tape capacity.

Additional tape drives have also been procured, and while they will clearly help with speed of data recovery in the event of a significant failure of the main archive, they will also be expected to support using the tape archive as an extension of the disk storage (since the expected volume of nearly all the categories of data in table \ref {tab:storage} is expected to exceed funding available rather quickly - however, such usage is not without problems (as outlined in the introduction), and so ideally data which has only primary copies on tape will be data that is expected to be rarely used. It is thus likely that JASMIN-CEMS disk storage capacity will need to grow again within 18 months or so, since frequent use data volumes are expected to exceed the available capacity on that time-scale.

\subsection{The JASMIN satellite nodes}

The other three JASMIN components also include compute resources based on the R610 platform, with Panasas deployed at Leeds and Reading as well. At Leeds and Reading the same basic configuration is being used as at RAL, but because Bristol already have a petascale storage associated with their university supercomputer, their JASMIN storage is integrated into the pre-existing university system.

In all the JASMIN sites the compute system is based on the same physical servers as that in the central JASMIN super-data-cluster. This will allow virtual machines to be constructed in the satellite sites that can then be migrated to the central system, and be expected to function with nearly the same performance at the central site - but with access to much more data. The discussion below on typical scientific workflows explains why JASMIN has been configured to support this capability.

\subsection{Wide area network environment}

Key links from the central JASMIN service are obviously to the university satellite nodes at Bristol, Leeds and Reading. These links use the JANET backbone, as do the links into the European Network for Earth Simulation (ENES) community, and the (global) Earth System Grid Federation.  Extensive network testing associated with support for CMIP5 have shown that when using GridFTP the JANET link can sustain  near wire-speed (disk-to-disk) to the west coast of the USA for hosts connected at 1 Gbit/s to the 10 Gbit/s network. The host upgrade to 10 Gbit/s is expected soon - but clearly it is not expected that wire-speed will then be available. Initial work with UPSCALE has shown that similar speeds can be sustained from PRACE's HERMIT to CEDA in production mode using multiple rsync streams (at the time of writing we were unable to deploy GridFTP at HERMIT). 

Despite those good figures for traffic, network contention is expected to be an issue for some key links, and where reliable bandwidth is required, JASMIN has commissioned, or will commission light paths.  The first light path in place is to the Royal Dutch Meteorological Institute, KNMI. KNMI will be using the light path to provide them with reliable access to the CMIP5 archive at BADC, both to support their scientific programme, and to support a new portal for the Climate Impacts community. Two further light paths will be commissioned during 2012 to link JASMIN to the two major climate community compute resources in the UK: the national supercomputing resource (HECToR) in Edinburgh, and the Met Office supercomputer (MONSooN) in Exeter. A link to DKRZ, the German Climate Compute centre will be provisioned over JANET using Tixel technology, so that we can ensure dedicated bandwidth at least in the last mile at both ends.

\begin{table}\centering

\begin{tabular}{lrrr} \toprule
Project & (TB) & JASMIN & CEMS \\ \midrule
Current BADC & & 350 & \\
Current NEODC & & & 300 \\
Current CMIP5 &  & 350 & \\
CEDA Expansion &  & 200 & 200 \\
CMIP5 Expansion &  & 800 & 300\\
CORDEX &  & 300 & \\
MONSooN Shared Data &  & 400 & \\
Other HPC Shared Data &  & 600 &   \\
User Scratch & & 500 & 300  \\ \midrule
Disk Totals & & 3500 & 1100\\ \bottomrule
\end{tabular}\caption{JASMIN-CEMS central disk storage requirements. (Only 1 PB of the existing CEDA 1.4 PB will be migrated to JASMIN-CEMS disk - the other 400 TB of of secondary data will not be migrated to spinning disk.) }
: \label{tab:storage}
\end{table} 

\section {JASMIN workflow}

This section will describe the workflow expected for the three main usages of JASMIN, but those explanations depend on a more detailed description of how compute and  virtualisation will be supported in the JASMIN and CEMS environments. Essentially JASMIN is providing three classes of service: a virtualised compute environment (not strictly a "private cloud", but we use that term anyway), a physical compute environment, and an HPC service ("Lotus"). The distinction between the latter two is  that the compute nodes do not have a private interconnect separate from the data interconnect, and could be easily reconfigured to join the private cloud (all the physical networking is in place), while the HPC nodes have their own network for interconnect, and could not be easily reconfigured into the JASMIN cloud. An additional hybrid cloud provides compute for the academic component of CEMS.

\begin{figure}[htp]\centering
\includegraphics[scale=0.65]{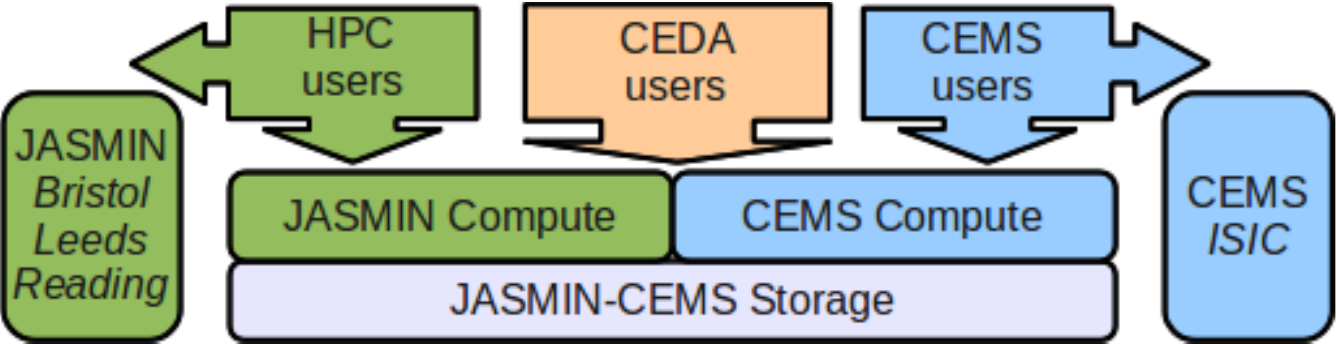}
\caption{Three broad classes of academic users are expected: those who need access to a high performance environment to support the analysis of simulation products ("HPC users"), many of whom will be able to exploit other JASMIN systems; the wider environmental science community who will ingest and access data and operational quality data services ("CEDA users"), and; researchers whose main interest flows from constructing and using observational data ("CEMS users"). Of course any individual may well use the system from any of these perspectives! (Commercial users will also be supported via the ISIC CEMS installation.)  \label{fig:disk-crop}}
\end{figure}

The clouds will initially be exploiting the vmWare virtualisation ecosystem, that is, the VM hosts will initially use the vSphere ESXi hypervisor to host virtual machines, which will be managed by vCentre server and vSphere. vCentre server provides high availability features for key machines, so that VMs running in physical hosts that suffer a hardware fault, can be either mirrored or automatically restarted on a different physical host. There will be two distinct clouds within the system as explained earlier, one to support manual VM provisioning by CEDA and the climate HPC community, and and one to support more dynamic provisioning by the academic users in the CEMS community. As a consequence, they have different vCentre server installations. 

The CEMS installation will form a vCloud director interface on top of its vCentre installation, separate from a similar installation in ISIC (a third cloud, but formally outside of JASMIN). The vCloud director will allow users to commission their own VMs onto a private network, which, though isolated from the site and the internet, will still able to access the Panasas storage.  By contrast, the manually commissioned VMs will be able to access the internet and have to abide to the security policies of the site networking. The CEMS VM machines, managed by vCloud director, will have no such remote access (although CEMS  will still be able to manually commission machines onto their cloud, which will be able to provide externally accessible websites etc).

Manually commissioned machines will achieve a higher level of trust primarily because it is possible to ensure extra levels of security via known customised machine configurations. e-Science have developed tools to run after basic virtual machines are created in the vSphere interface (which configures basic properties such as the size of the  system image disk, memory, and number of cpus). These tools will initially deploy a basic Linux system (initially RHEL5, RHEL6, SL5 or SL6) with the latest OS patches, and configured to remotely log key properties and processes.

\subsection {Supporting CEDA}

The CEDA compute environment has three basic requirements: to support data ingestion, to support data management (inspection, documentation, visualisation, processing, migration, replication), and to support data extraction and download. 

At the time of writing (early 2012), CEDA hosts 1.4 PB of data on disk with an additional 1.3 PB on tape. Primary data copies are stored on Network Attached Storage (NAS) with extra copies on NAS and/or tape depending on the status of the datasets involved. (Datasets for which CEDA is "the archive of last resort" typically have three or four copies in the archive, datasets for which other copies can be retrieved, generally have fewer or no copies in the archive). Approximately 90 storage servers provided over 340 storage pools supporting 540 distinct sets of files. User and management computing deployed roughly 200 physical and virtual computers. Virtualisation used the Xen hypervisor system, and was mainly used to provide clean environments around specific web services. The systems were distributed across two machine rooms, connected by a complex environment of routers and switches.  The entire edifice was maintained without any sophisticated management tools, having grown organically over the previous decade from what was then a relatively simple system based around storage attached to one or two large servers. No financial headroom had ever been made for any other solutions!  Not surprisingly, efficiency was falling - not only from a system admin point of view, but from a data management point of view - nearly all staff were spending significant amounts of time migrating data between storage pools, and some network performance problems had remained undiagnosed for more than a year.  (Despite that, in the previous full reporting year, user satisfaction was high, and over 3,000 users had downloaded over 200 TB of data in over 20 million files.) Nonetheless, clearly CEDA could not continue to both grow and maintain user satisfaction from that position - yet the projection for the next year suggested a doubling in archive size.

JASMIN was conceived of primarily as a solution to these problems - a computing environment more appropriate to a petascale environment, more scalable, and providing the basic infrastructure to support both archival services, and data usage. However, the CEDA mission is both to curate the data products from, and facilitate the science programmes of, their communities. To that end, safe scalable archival is not enough - if that were the only goal, then most of the CEDA problems listed above would simply be pushed down into the university departments engaged in "big data science" who would all individually have to deal with petascale data, high performance networks etc. While a degree of that is unavoidable, the JASMIN solution was also designed to provide a national data analysis service for the climate, earth observation, and earth system science community. Hence, JASMIN has been configured to provide high performance data access and analysis, as well as reliable data archival with a low total cost of ownership.

During 2012, all the CEDA data and services will migrate onto JASMIN.   As discussed earlier, some of the physical layout of the storage is constrained by the necessity to ensure archival fidelity - with both logical and file system separation between the formal archive and more transient data. The existing plethora of virtual machines (mainly supporting extraction, processing and visualisation) will be consolidated onto the JASMIN data compute. Basic web portals and management functions (such as ingestion and metadata creation)  will be carried out on the dedicated management compute system.   External CEDA users requiring direct access to  the data will do so through either their own virtual machines, or through a shared virtual machine configured for community login access - however it is expected that most CEDA users will interact through web services and portals - all of which should be easier to manage and perform more reliably on JASMIN.  

\subsection {Parallel Data Analysis}

The advent of large-scale data and the consequential analysis problems have led to two new challenges for the research community: how to share such data to get the maximum value and how to carry out efficient analysis. (One might argue that sharing data has always been desirable, here we are arguing that is now {\it necessary}, since no one individual or even small research group, is or are likely to have the relevant expertise to fully analyse the complexity of some modern data sets). Solving both challenges require a form of parallelisation: the first is social parallelisation (involving trust and information sharing), the second data parallelisation (involving new algorithms and tools).

JASMIN aims to support both forms of parallelisation: social parallelisation will be supported by the use of "project workspaces", and technical parallelisation by the cloud computing discussed earlier. By allowing HPC data to be downloaded into "project workspace", communities will be empowered to grant access to prospective analysts. Even JASMIN however, has limited storage resource, so a policy driven environment will govern what data can remain on project disk or in the JASMIN project space tape store.  This policy will need to take a holistic view of the storage available across the UK - covering all the JASMIN locations, the large scale archive available at HECToR, and additional tape systems at the Met Office. Data will be expected to be migrated between locations according to the policy. Initially the policy environment will be managed manually, but obviously it would be desirable to automate it when effort to develop and/or implement such systems is available. iRODS \cite{moore_management_2011} is obviously one candidate technology, but some experimentation would be necessary before it could be used operationally. In the mean time, data migration will use bespoke software procured as part of JASMIN.  However, whatever system is in place, this policy will require information to support decisions, so datasets stored in project workspaces will require some minimal metadata. 

Technical parallelisation requires developing algorithms which can exploit the cluster environment on JASMIN. Analysis parallelisation is in its infancy in climate sciences, with few tools capable of exploiting any parallel environment - beyond manual scripting of the use of multiple processors. In fact, the only published descriptions of significant parallelisation capability are for the NetCDF operators (NCO, see \url{http://nco.sourceforge.net}), which can support OpenMP or MPI as appropriate for most operators \cite{zender_high_2007}. Some OpenMP parallelisation is apparently\footnote{\url{https://code.zmaw.de/projects/cdo/wiki/OpenMP_support}, accessed 13 March 2012}  also possible with the Climate Data Operators (CDO, see \url{http://code.zmw.de/projects/cdo}) package when compiled appropriately.  Other packages are beginning to support native parallelisation as well, and there has been some initial work with MapReduce algorithms (e.g. \cite{buck_scihadoop:_2011}, \cite{wang_scimate:_2012}).

Operator based packages like NCO and CDO are intended to support scripting of calculations on files, but the resultant scripts can be very large (e.g. \cite{zender_high_2007} describe a 14,000 line script carrying out a 
relatively simple extraction and co-location activity).  To get shorter, more maintainable and evolvable codes, many scientists prefer to do much of their data manipulation with higher level languages, with IDL, Matlab, and latterly, python, being popular. However, we are not aware of any libraries in any of these languages which provide easy access to parallelisation for typical data manipulation tasks involving climate or earth observation data. Hence, we believe the community will first exploit JASMIN parallelisation by scripting the use of codes written in these languages to exploit the easy parallelisation available when carrying out data analysis tasks which are repeated for many timesteps. (In contrast to simulation, in data analysis, many tasks yield easy parallelisation over time.) For all these reasons we expect such users are initially most likely to want to port their existing analysis environments into JASMIN - this is the reason why JASMIN supports virtual machine creation as a key platform for user services.

We anticipate that users in Leeds, Bristol and Reading might well begin by developing virtual machines in their own instances of JASMIN, and then migrate such virtual machines to the central JASMIN resource when they need access to larger datasets or more compute.  This new arrangement will be analogous to the migration of regular HPC jobs from a university system to a national compute system when bigger jobs or faster time to solution are required.

In the medium term, CEDA will work with the community to develop (or modify existing packages) to provide "out-of-the-box" support for parallelisation by coupling tools such as ipython (\url{http://ipython.org/})and cf-python (\url{http://code.google.com/p/cf-python/}).

\subsection{Exploiting HPC-type parallelisation}

The "Lotus" HPC cluster supports two more significant JASMIN use cases, the use of parallel input output libraries in earth system models, and in data analysis codes.  Neither are particularly innovative concepts - parallelisation is in use in data analysis workflows, e.g. \cite{zender_high_2007}, and a parallel NetCDF library \cite{li_parallel_2003} is in use in some models \cite{dennis_application-level_2011}, but it is not in use in typical UK workflows and models.

Parallel input/output is a big part of reducing execution times for complex jobs, and is a particular limiter for earth system models as they are required to scale onto massively parallel systems \cite{dennis_application-level_2011}. While Lotus is clearly not a massively parallel system, it will provide a platform on which parallel i/o systems can be tested in the context of both earth system models and data analysis systems.  The latter will be an important part of gaining strong parallelisation (beyond the weak parallelisation gained via decomposing analysis along time slices) in data analysis. (It is worth noting that this is another area where a truly parallel I/O file system like Panasas gains over the previous CEDA environment - parallel I/O using MPIO would gain no advantage on a shared NFS file system.) 

Lotus will also be used as a generic testing system, allowing those responsible for model development and testing faster turn around than can be gained on the national HPC systems. Such testing could include some limited use of the data archive to support data assimilation (limited, because it is not intended that the data archives at JASMIN include the full suite of meteorological observations used in, for example, numerical weather prediction assimilation systems) Any unused cycles will be used for running lower resolution earth system models (ones that do not make effective use of the greater parallelisation available elsewhere).

\subsection {Supporting interfaces and services}

The previous subsections described fundamental data archives and data analysis using systems designed and managed by those "in" the community.  The JASMIN-CEMS system is also designed to support a more service orientated interface, going well beyond the "software as a service" or "web-service" interfaces provided by CEDA.

CEMS will be able to support allocating  different research groups, universities, even companies, slices of the total cloud resource for their own purposes - exploiting the multi-tenancy capability provided by vCloud director to allocate reserved pieces of the total cloud resource to deliver "Infrastructure-as-a-service" (IaaS).  These resources could involve customised processing environments, targeted at specific algorithm development or even delivering specific services. The flexible environment would support simple set-up and tear down, allowing testing both of scientific algorithms and commercial services, without significant upfront investment.  The ISIC partners are also likely to deploy platform-as-a-service (PaaS) since it is attractive in the commercial environment, allowing much more constraints and control (over what clients can do, what licenses are needed and what data can be accessed) than the more unconstrained offering of "empty" virtual machines inherent in IaaS.

It is expected that as users experiment in the CEMS environment they could find that these algorithms become something generically useful to the wider community.  They could then become part of an 'official' CEMS library of processing algorithms and be deployed via Software-as-a-service (SaaS) behind portals and/or web-service interfaces.

\section {Summary}

Continuing advances in both computing and remote sensing are leading to ever increasing volumes of data from simulations and observations of the environment. Using these data efficiently, on their own, and together, is becoming more and more challenging, and these challenges are being distributed not only amongst existing scientific communities who at least have some familiarity with the territory, but new communities of scientific and policy users who need to make sense of data which is generally not organised well for their problems.

CEDA is charged both with managing such data for posterity, and facilitating the efficient delivery of the UK environmental science programme. The existing CEDA computing environment (hundreds of millions of files organised into hundreds of "filesets" distributed over hundreds of disk partitions on dozens of servers) is no longer fit for purpose. Managing the data in that environment had become a challenge, let alone facilitating their use. 

JASMIN and CEMS are part of the solution to these problems: delivering components of an integrated national storage strategy - linking environmentally relevant storage (the EPSRC tertiary storage system and components of JASMIN and CEMS)  and compute (HECToR, MONSooN, and JASMIN and CEMS themselves). JASMIN and CEMS are providing computing resources (configured in a range of ways) coupled to petascale   storage systems with an initial capacity of 9.4 PB which include very high performance reliable disk (4.6 PB usable, 6.6 PB raw) and concomitant tape resources (initially 4.8 PB, but scalable).

The systems will be used not only to support the CEDA data centre functions, but to directly support the scientific community, allowing the development of parallel data analysis tools which can be exploited both on relatively small numbers of cores typical of customised virtual machines, and order dozens to hundreds of cores using clusters configured for HPC.  The virtualisation environment will help researchers develop and utilise their analysis codes using their own environments, thus increasing productivity, as well as aid in migration to more scalable solutions in the larger JASMIN system. The virtualisation environment will also support the more efficient deployment and scalability of existing data services (subsetting, visualisation, download) as well as the construction of new services including customisable workflows developed and deployed by third parties - both academic (using JASMIN and the academic component of CEMS described here) and business (using the ISIC component of CEMS which has not been described here).

\printbibliography

\end{document}